\documentclass[journal]{IEEEtran}

\ifCLASSINFOpdf

\else

\fi
\usepackage{epsfig} % for postscript graphics files
\usepackage{subfigure}
\usepackage{graphicx} % for pdf, bitmapped graphics files
\usepackage{epstopdf}
\usepackage{mathptmx} % assumes new font selection scheme installed
\usepackage{times} % assumes new font selection scheme installed
\usepackage{amsmath} % assumes amsmath package installed
\usepackage{amssymb}  % assumes amsmath package installed
\usepackage{algorithmic}
\usepackage{algorithm}
\usepackage{textcomp}
\usepackage{xcolor}
\usepackage{cases}
\usepackage{float}
\newtheorem{theorem}{Theorem}
\newtheorem{lem}{Lemma}

\newtheorem{remark}{Remark}
\newtheorem{assump}{Assumption}
\begin{document}

\title{Adaptive Finite-time Disturbance Rejection for Nonlinear Systems using an Experience-Replay based Disturbance Observer}

\author{Zhitao~Li, Amin~Vahidi-Moghaddam,   Hamidreza~Modares ~and~Jinsheng~Sun
% <-this % stops a space
\thanks{Z.~Li and J.~Sun are with School of Automation, NanJing University Science $\&$ Technology, NanJing 210094, China. (e-mail:jssun67@163.com)}
\thanks{H.~Modares and A.~Vahidi-Moghaddam are with department of Mechanical Engineering, Michigan State University,
East Lansing, MI, 48863, USA. }
%\thanks{L.~Gao is with Institute of Intelligent Systems and Decision, Wenzhou University, Zhejiang 325035, China.}
}

\markboth{Journal of \LaTeX\ Class Files,~Vol.~X, No.~X, August~201X}%
{Shell \MakeLowercase{\textit{et al.}}: IEEE Journals}
\maketitle

\begin{abstract}
\textcolor{black}{Control systems are inevitably affected by external disturbances, and a major objective of the control design is to attenuate or eliminate their adverse effects on the system performance.} \textcolor{black} {This paper presents a disturbance rejection approach with two main improvements over existing results: 1) it relaxes the requirement of calculating or measuring the state derivatives, which are not available for measurement, and their calculation is corrupted by noise, and 2) it achieves finite-time disturbance rejection and control. To this end, the disturbance is first modeled by an unknown dynamics,}  and an  adaptive disturbance observer is proposed to estimate it. A filtered regressor form is leveraged to model the nonlinear system and the unknown disturbance. It is shown that using this filtered regressor form, the disturbance is estimated using only measured state of the \textcolor{black}{ regressor. That is, contrary to the existing results on disturbance rejection, the  presented approach does not require the state derivative measurements.}  \textcolor{black}{To improve the convergence speed of the disturbance estimation, an adaptive law, equipped with experience replay, is presented. The disturbance observer is then augmented with an adaptive integral terminal sliding mode control to assure the finite-time convergence of tracking error to zero.} A verifiable rank condition on the history of the past experience used by the experience-replay technique provides a sufficient condition for convergence. Compared  to the existing results, neither the knowledge of the disturbance dynamics nor the state derivatives are required, and finite-time stability is guaranteed. A simulation example illustrates the effectiveness of the proposed approach.

\end{abstract}


\section{Introduction}
 {\color{black}Disturbances can be inevitably found in almost every control system and, if not rejected, they can drastically jeopardize the system's performance. Therefore, it has been a long stand challenge to reject disturbances in control society.} Existence of persistent disturbances is one of the sources of difficulties in achieving a good system performance in applications such as {\color{black} marine vessels  \cite{hu2019adaptive}, active vibration suppression \cite{zhang2019prescribed}, tracking of a reference position \cite{hu2019adaptive, cabecinhas2014nonlinear}, and rotating mechanisms control \cite{gentili2003robust}.} Disturbances are not measurable in most real-world applications, but have some structures, possibly unknown, which must be leveraged by the control design to achieve a better performance. For instance, the disturbance in surprisingly large number of applications can be reasonably modeled as the output of a dynamical system, called exosystem, with unknown dynamics. For example, in systems with rotating, the disturbance source often consists of many of periodic components with unknown frequencies (e.g. engine noise in automobile and aircraft). {\color{black}  Modeling the disturbance with an exosystem dynamics  is a standard practice and has been considered by many researchers \cite{ bacsturk2013adaptive, basturk2014state, madonski2020active, yilmaz2019adaptive}.}

The most common approach for disturbance cancellation is the internal model principle \cite{pan2016internal} for which the disturbance dynamics is incorporated into the controller design. A related problem is the output regulation \cite{zhao2016semi} for which the system is supposed to track a reference trajectory and/or reject a disturbance with known exosystems. If the dynamics of the exosystem generating the disturbance is known, and the disturbance can be measured, these approaches can be directly used to completely  reject  the disturbance. However, in reality, neither the exosystem dynamics is known, nor {\color{black} can we measure the disturbance. In \cite{yilmaz2019adaptive}, an adaptive output feedback scheme with  adaptive backstepping is presented  to reject the disturbances by assuming that the state derivatives are measurable.} In \cite{sun2016distributed, wang2019consensus},  disturbance observers have been designed  for the case where the disturbance cannot be measured, but the exosystem dynamics is assumed  to be  known.  To relax the requirement of knowing the exosystem dynamics, {\color{black} adaptive state-derivative feedback techniques have been presented for } both matched disturbances \cite{bacsturk2013adaptive} and mismatched disturbances \cite{basturk2014state, basturk2018active}. {\color{black}However, to estimate the disturbance, the state derivatives are assumed {\color{black} to be available which  usually cannot be  directly sensed and must be  calculated from the consecutive state measurement, which is corrupted by noise.} } Moreover, the history of  the interaction between the disturbance and the system is not taken into account in the existing results to achieve better convergence and consequently improve the system's  performance.

Finite-time stability has attracted a surge of interest in both model-based and model-free control due to its desired properties. {\color{black} Specifically, variants of sliding mode control (SMC) \cite{slotine1991applied, khalil2002nonlinear}, such as terminal sliding mode control (TSMC) \cite{li2019position, wang2019practical} have been presented to guarantee the finite-time stability}. Moreover, integral TSMC (ITSMC) \cite{qiao2019trajectory} {\color{black} has been successfully used to achieve the finite-time stability and solve the singularity problem in TSMC.} Successful  applications of variants of ITSMC for robot manipulators  \cite{rahmani2016hybrid} and autonomous underwater vehicles \cite{qiao2017adaptive} have also been reported. Disturbance rejection control  has also been studied using SMC and adaptive TSMC in  \cite{yang2013continuous, feng2019integral, rabiee2019continuous}. However,  to achieve finite-time stability, the worst-case bound of the disturbance is considered in the design, which results in unnecessary large control efforts and excessively conservative controllers. To obviate this issue, the structure of the disturbance can be leveraged to estimate it and provide the controller with quantified and decaying disturbance bounds. This significantly improves the performance of the controller.

In this paper, we present a novel adaptive finite-time disturbance rejection controller that  {\color{black} does not require  the knowledge of the disturbance dynamics and the state derivatives. Towards this goal, we first introduce a new adaptive disturbance  observer by formulating its dynamics into a filtered regressor form to overcome the shortcoming of requiring the state derivative measurements which are not usually available and their calculation is corrupted by noise.}  Then, we design an observer to estimate the unknown disturbance and its dynamics. Next, we present a novel experience replay-based adaptive disturbance observer, in which the history of the data collected along the system  trajectories {\color{black}is incorporated into the update law} to guarantee the exponential convergence of the disturbance estimation error under satisfying a rank condition on the history stack. This is inspired by how declarative memory (explicit memories that can be inspected and recalled consciously) in human brain stores data to reduce the number of interactions with the environment to learn it. We show that reusing {\color{black} the experiences} increases the efficiency of data-based disturbance estimation. Finally, the disturbance {\color{black} observer} is augmented  with an adaptive ITSMC assuring {\color{black} that the} tracking error goes to zero in finite time. The adaptive controller's  gain  follows the variation tendency of the disturbance to avoid overestimating the disturbance. {\color{black} This is less control-energy demanding than the existing adaptive ITSMC results for disturbance rejection as they have been designed based on the maximum disturbance bound.}  A simulation is finally provided to verify the effectiveness of the proposed approach.

Notations: In this paper, $\mathbb{R}^{n}$ and  $\mathbb{R}^{n \times m}$ represent a real  $ n-$ dimensional vector and a real $n \times m$ matrix, respectively.  For a matrix $A$, $A^{T}$ stands for {\color{black}its  transpose}, $A^{+}$ {\color{black}stands for its  generalized inverse, and if} matrix $A$ has full row rank (or column rank) $A^{+}= A^{T}(AA^{T})^{-1}$ (or $A^{+}= (A^{T}A)^{-1}A^{T})$ {\color{black}stands for its pseudoinverse.}
$A_{vec}$ stacks the columns of the matrix $A$ into a vector. $\lambda_{min}(A)$  and $\lambda_{max}(A)$ {\color{black} represent the minimum and maximum eigenvalues  of $A$.  Moreover $(A \otimes B)$ represent the Kronecher product of A and B.  }The function $f(t)$  belongs to $L_{2}$ and $L_{ \infty}$ spaces, i.e, $f(t) \in L_{2}$ and $f(t) \in L_{\infty}$, if it satisfies {\color{black} $\int_{\infty}^{0}f(t)^{T}f(t)dt< \infty$ } and $sup_{t \in R}|f(t)| < \infty$, respectively.

\section{A filtered regressor form for modeling the system and the disturbance dynamics}
In this section, a nonlinear dynamical system with unknown disturbance is introduced. Then, a filtered regressor form is employed to model the nonlinear system dynamics and the disturbance exosystem dynamics.

Consider the following nonlinear dynamical systems
\begin{equation}\label{nonlinear systems}
\dot{x}=f(x)+g(x) u(x)+D \varepsilon_{T}
\end{equation}
 where $x \in \mathbb{R}^{n}$ is a measurable system state vector, $f(x) \in \mathbb{R}^{n}$ is the drift dynamics of the system, $g(x) \in  \mathbb{R}^{n \times m} $  is the input dynamics of the system and assumed to be full  column rank, and $u(x) \in \mathbb{R}^{m}$ is the control input. Moreover, $D \in  \mathbb{R}^{n \times d}$ is the disturbance dynamics, and $\varepsilon_{T} \in \mathbb{R}^{d}$ is the disturbance. It is assumed that the unknown disturbance $\varepsilon_{T}$ is generated by the following dynamics
  \begin{equation}\label{disturbance dynamic}
   \dot{\varepsilon}_{T}=S \varepsilon_{T}
 \end{equation}
 where $S \in \mathbb{R}^{d \times d}$ is an unknown matrix of appropriate dimension.

 \begin{assump}\label{assump1}
  The system (\ref{nonlinear systems}) is stabilizable. Moreover, $f(0)=0$,  and $f(x)$ and $g(x) $ are locally Lipschitz.
 \end{assump}

 \begin{assump}\label{assump2}
  {\color{black}The matrix} $S$ is unknown with eigenvalues on the imaginary axis.
 \end{assump}
{\color{black}
\begin{remark}\label{disturbance dynamic remark}
Note that under Assumption \ref{assump2},  the disturbance dynamics (\ref{disturbance dynamic})  can generate external sinusoidal disturbances and many other periodic disturbances that are common in many practical applications \cite{hu2019adaptive, bacsturk2013adaptive, basturk2014state}. {\color{black} Moreover,  if the eigenvalues of S are located in the left-half side of the imaginary axis, it results in a temporary disturbance that its affects will go away and can be ignored. On the other hand, the eigenvalues of S cannot be in the right-hand side since it indicates an unstable exosystem;  thus,  the disturbance will be unbounded with infinite energy, which is not realistic.}

\end{remark}}

We now present a filtered regressor form of the system dynamics (\ref{nonlinear systems}) and the disturbance dynamics (\ref{disturbance dynamic}).

 Let the functions $f(x)$ and $g(x)$  be  parameterized as
 \begin{equation}\label{NN approximatefg}
   f(x)=\theta^{*} \xi(x)  \ \ \ g(x)=\psi^{*}\zeta(x)
 \end{equation}
where $\theta^{*} \in \mathbb{R}^{n \times p_{\theta}}$ and $\psi^{*} \in \mathbb{R}^{n \times p_{\psi}}$ are the known weights matrices, $\xi(x) \in \mathbb{R}^{p_{\theta}}$ and $\zeta(x) \in \mathbb{R}^{p_{\psi} \times n}$ {\color{black} are the known basis functions,}  $p_{\theta}$ and $p_{\psi}$ are the dimensions of the system dynamics $f(x)$ and $g(x)$. Note that since $f(x)$ and $g(x)$ are known, $\theta^{*}$, $\psi^{*}$, $\xi(x)$ and $\zeta(x)$ can always be founded and also assured known.
Then, from (\ref{NN approximatefg}), the system (\ref{nonlinear systems}) can be written as
  \begin{equation}\label{reweitten nonlinear systems}
    \dot{x}=\theta^{*} \xi(x)+\psi^{*}\zeta(x)u(x)+D\varepsilon_{T}
  \end{equation}
or equivalently
\begin{equation}\label{re nonloear system}
 \dot{x}=\phi^{*} z(x, u)+D\varepsilon_{T}
\end{equation}
 {\color{black}where $\phi^{*} \in \mathbb{R}^{n \times d}$ is the known weights matrix, and} $z(x,u)=[\xi^{T}(x)\ \ u(x)^{T} \zeta^{T}(x)] \in \mathbb{R}^{d}$ is the regressor vector.

  Inspired by \cite{modares2013adaptive}, the  filtered regressor forms of the  system (\ref{nonlinear systems}), (\ref{re nonloear system}) and the disturbance {\color{black}dynamics (\ref{disturbance dynamic}) are}
 given by Lemma \ref{lem1} and Lemma \ref{lem2}, respectively.
 \vbox{}
\begin{lem}\label{lem1}
The system (\ref{nonlinear systems}), (\ref{re nonloear system}) can be expressed as
\begin{equation}\label{filtered regressor}
\begin{array}{l}
{x=\phi^{*} h(x)+a l(x)+\bar{\varepsilon}+\rho(t)}, \\
{\dot{h}(x)=-a h(x)+z(x, u)}, h(0)=0,\\
{\dot{l}(x)=-a l(x)+x}, l(0)=0,\\
{\dot{\rho}(t)=-a\rho(t)},\rho(0)=x(0),\\
 {\dot{\varepsilon}=
-a\varepsilon+\varepsilon_{T}}, \varepsilon(0)=0\\
{\bar{\varepsilon}=D\varepsilon},
\end{array}
\end{equation}
 where $a>0$, $h(x) \in \mathbb{R}^{d}$ is the filtered regressor version of $z(x, u)$,  $l(x)\in \mathbb{R}^{n}$ is the filtered regressor version of the state $x$,  $\varepsilon$ is the filtered  disturbance state,   $\bar{\varepsilon}$ is the filtered output disturbance .
\end{lem}

\begin{IEEEproof}
Adding and subtracting  the  term $ax$ with $a>0$  to  (\ref{re nonloear system}),  one has
 \begin{equation}\label{auxiliary system}
  \dot{x}=-ax+\phi^{*}z(x, u)+ax+D\varepsilon_{T}
 \end{equation}
 or equivalently
\begin{equation}\label{rewitten x}
  \dot{x}_{i}=-a x_{i}+\phi_{i}^{*} z(x, u)+a x_{i}+(D \varepsilon_{T})_{i},\ \ i=1, \ldots, n
\end{equation}
where $\phi_{i}^{*}$ and $(D \varepsilon_{T})_{i}$  are the $i-th$  rows of the weights matrix $\phi^{*}$ and disturbance $D\varepsilon_{T}$, respectively.

Thus, the solution of (\ref{rewitten x}) can be expressed as
\begin{equation}\label{solution filiter}
\begin{split}
  x_{i}(t)=&e^{-a t} x_{i}(0)+\int_{0}^{t} e^{-a(t-\tau)} \phi_{i}^{*} z(x, u) d \tau
\\ & +a \int_{0}^{t} e^{-a(t-\tau)} x_{i}(\tau) d \tau + \int_{0}^{t} e^{-a(t-\tau)}(D \varepsilon_{T})_{i} d \tau
\end{split}
\end{equation}
 Define
 \begin{equation}\label{h filter}
   h(x)=\int_{0}^{t} e^{-a(t-\tau)} z(x, u) d \tau
 \end{equation}
 \begin{equation}\label{varepsilonbar filter}
   \bar{\varepsilon}_{i}(x)=\int_{0}^{t} e^{-a(t-\tau)}(D \varepsilon_{T})_{i} d \tau, \ \ i=1, \ldots, n
 \end{equation}
 \begin{equation}\label{l filter}
   l_{i}(x)=\int_{0}^{t} e^{-a(t-\tau)} x_{i}(\tau) d \tau, \ \ i=1, \ldots, n
 \end{equation}
 \begin{equation}\label{rho filter}
   \rho_{i}(t)=e^{-a t} x_{i}(0),\ \ i=1, \ldots, n
 \end{equation}
 \begin{equation}\label{varepsilon filter}
  \varepsilon_{i}=\int_{0}^{t} e^{-a(t-\tau)} \varepsilon_{T_{i}} d \tau, \ \ i=1, \ldots, n
 \end{equation}
  Then,  using (\ref{h filter})-(\ref{varepsilon filter}), (\ref{solution filiter}) becomes
\begin{equation}\label{solution filiter change}
  x_{i}=\phi_{i}^{*} h(x)+a l_{i}(x)+\bar{\varepsilon}_{i}+\rho_{i}(t)
\end{equation}
Let $l(x)=[l_{1}(x), l_{2}(x), \ldots, l_{n}(x)]^{T}$, $\varepsilon=[\varepsilon_{1}, \varepsilon_{2}, \ldots, \varepsilon_{n}]^{T}$,
$\bar{\varepsilon}=[\bar{\varepsilon}_{1}, \bar{\varepsilon}_{2}, \ldots, \bar{\varepsilon}_{n}]^{T}$, and $\rho(t)=[\rho_{1}(t), \rho_{2}(t), \ldots, \rho_{n}(t)]^{T}$.
The matrix form of (\ref{solution filiter change}) can be written as
\begin{equation}\label{solution filiter change form}
  x=\phi^{*} h(x)+a l(x)+\bar{\varepsilon}+\rho(t)
\end{equation}
On the other hand, using (\ref{varepsilonbar filter}) and
(\ref{varepsilon filter}), one has
\begin{equation}\label{disturance with D}
  \bar{\varepsilon}=D\varepsilon
\end{equation}
Taking derivative of $h(x)$, $l(x)$,  and $\varepsilon$ results in
 \begin{equation}\label{derivation h}
   \begin{split}
    \dot{h}(x)=&-a\int_{0}^{t} e^{-a(t-\tau)}z(x,u)d\tau + (z(x,u)-0)
    \\=& -a h(x)+ z(x,u)
    \end{split}
 \end{equation}
 \begin{equation}\label{derivation l}
   \begin{split}
    \dot{l}(x)=&-a\int_{0}^{t} e^{-a(t-\tau)}x(\tau) d\tau + (x-0)
    \\=& -a l(x)+ x
    \end{split}
 \end{equation}
\begin{equation}\label{derivation varepsilon form}
\begin{split}
  \dot{\varepsilon}=&-a \int_{0}^{t} e^{-a(t-\tau)} \varepsilon_{T} d \tau +\varepsilon_{T}\\
  =& -a\varepsilon+\varepsilon_{T}
 \end{split}
\end{equation}
Note that $\rho(t)= e^{-at}\rho(0)$ gives $\dot{\rho}(t)= -a\rho(t)$ with $\rho(0)=x(0)$.
This completes  the proof.
\end{IEEEproof}

Similarly to Lemma \ref{lem1}, a filtered regressor form for the unknown disturbance dynamics (\ref{disturbance dynamic}) are shown as the following Lemma

\begin{lem}\label{lem2}
 The disturbance dynamics (\ref{disturbance dynamic}) can be expressed as

\begin{equation}\label{filtered disrurbance}
\begin{array}{l}
{\varepsilon_{T} =(S+aI_{d})\varepsilon}+\rho_{\Delta}(t),\\
{ \dot{\varepsilon}=-a\varepsilon+ \varepsilon_{T}}, \varepsilon(0)=0,\\
{\dot{\rho}_{\Delta}(t)=-a\rho_{\Delta}(t)}, \rho_{\Delta}(0)=\varepsilon_{T}(0)
\end{array}
\end{equation}
where $a>0$ is a constant, and $\varepsilon=\int_{0}^{t} e^{-a(t-\tau)}\varepsilon_{T}(\tau) d \tau $.
\end{lem}

\begin{IEEEproof}
Adding and subtracting the term $a\varepsilon_{T}$ with $a>0$  to the right-hand side of the system (\ref{disturbance dynamic}), one has
\begin{equation}\label{auxiliary system disturbance}
  \dot{\varepsilon}_{T}=-a\varepsilon_{T}+S\varepsilon_{T}+
  a\varepsilon_{T}
 \end{equation}
The solution of (\ref{auxiliary system disturbance}) can be written as
\begin{equation}\label{solution filiter disturbance}
\begin{split}
  \varepsilon_{T}=& e^{-at}\varepsilon_{T}(0)+S\int_{0}^{t} e^{-a(t-\tau)}\varepsilon_{T}(\tau) d \tau
 \\&+a \int_{0}^{t} e^{-a(t-\tau)}\varepsilon_{T}(\tau) d \tau
\end{split}
\end{equation}
Defining $\varepsilon=\int_{0}^{t}e^{-a(t-\tau)}\varepsilon_{T}(\tau) d \tau$ and $\rho_{\Delta}(t)=e^{-at}\varepsilon_{T}(0)$, (\ref{solution filiter disturbance}) becomes the first equation in (\ref{filtered disrurbance}). On the other hand, the derivative of filtered  disturbance state $\varepsilon$ becomes the second equation in (\ref{filtered disrurbance}), and the derivative of  $\rho_{\Delta}(t)$ becomes the third equation in (\ref{filtered disrurbance}).
This completes the proof.
%is the same as (\ref{varepsilon form}),  $\varepsilon$  the same as  Lemma \ref{lem1}. Then we get Lemma \ref{lem2}
%directly, so we omit it.
\end{IEEEproof}

\begin{remark}\label{rem1}
Note that the filtered  disturbance state $\varepsilon$ in the filtered regressor form of the disturbance dynamics (\ref{filtered disrurbance}) is the same as  $\varepsilon$ in the filtered regressor form of the system dynamics (\ref{filtered regressor}). On the other hand, based on (\ref{filtered regressor}), $\bar{\varepsilon}=D\varepsilon=x-\phi^{*} h(x)-a l(x)-\rho(t)$ which can be measured since it only depends on the state $x$. That is,  $\varepsilon=D^{+}\bar{\varepsilon}$ can be calculated using only the state measurements. Therefore, to estimate the disturbance in (\ref{filtered disrurbance}), we only need to estimate the unknown dynamic matrix $S$. An observer is designed next to estimate $S$. This is in contrast to the existing disturbance estimation results that require measurements of the state derivatives as well \cite{bacsturk2013adaptive, basturk2014state}.
\end{remark}

\section{An adaptive disturbance observer using measured system's states}
Since the disturbance $\varepsilon_{T}$ cannot be
 measured and only the system's state is assumed to  be measurable, we design a disturbance observer using the filtered regressor form (\ref{filtered disrurbance}) as follows
 \begin{equation}\label{observer disturbance}
 \begin{array}{l}
  \hat{\varepsilon}_{T}=(\hat{S}+aI_{d})\varepsilon+
  \hat{\rho}_{\Delta}(t) \\
  \dot{\hat{\rho}}_{\Delta}(t)=-a\hat{\rho}_{\Delta}(t),
  \hat{\rho}_{\Delta}(0)=\hat{\varepsilon}_{T}(0)
   \end{array}
 \end{equation}
where $\hat{S}$ is the estimation of the disturbance weights matrix $S$. Note that as stated in Remark \ref{rem1}, $\varepsilon$ is measured using only the measured states.

To design an adaptive disturbance observer, the following auxiliary dynamics are used to develop an adaptive law for $\hat{S}$.

\begin{equation}\label{estimate state}
\begin{aligned}
  \hat{x}&=\phi^{*} h(x)+a l(x)+\hat{\bar{\varepsilon}}+\rho(t)\\
  \dot{\hat{\varepsilon}}&=-a\hat{\varepsilon}+
  \hat{\varepsilon}_{T} \\
  \hat{\bar{\varepsilon}}&=D\hat{\varepsilon}
  \end{aligned}
\end{equation}
{\color{black}where $\hat{x}$ is an auxiliary variable used for estimation of the disturbance}, $\hat{\varepsilon}$ is the estimated filtered disturbance, and  $\hat{\bar{\varepsilon}}$ is the estimated filtered output disturbance.
{\color{black}
\begin{remark} \label{explain hat x}
In this paper,  the state $x$  {\color{black} is assumed to be available for measurement. Note that in (\ref{estimate state}), $\hat{x}$  is not actually the state estimation and is only used to measure the disturbance.}
\end{remark}}
 Defining $\tilde{\varepsilon}_{T}=\varepsilon_{T}-
\hat{\varepsilon}_{T}$,  $\tilde{S}=S-\hat{S}$, and $\tilde{\rho}_{\Delta}=\rho_{\Delta}-\hat{\rho}_{\Delta} $, and using  (\ref{filtered disrurbance}), (\ref{observer disturbance}), one has

\begin{equation}\label{error observer}
\begin{split}
 \tilde{\varepsilon}_{T}=&(S+aI_{d})\varepsilon+
 \rho_{\Delta}-
 (\hat{S}+aI_{d})
 \varepsilon -\hat{\rho}_{\Delta}\\=&\tilde{S}\varepsilon
 +\tilde{\rho}_{\Delta}
 \end{split}
\end{equation}

Defining  $\tilde{e}=x-\hat{x}$,  the adaptive law for $\hat{S}$  is designed as

\begin{equation}\label{adaptive law}
  \dot{\hat{S}}_{vec}=\Gamma(\bar{\varepsilon}^{T}F^{T} \otimes D)^{T}\tilde{e}
\end{equation}
where $\hat{S}_{vec}$ is the estimated vector obtained by stacking rows of the unknown matrix $\hat{S}$,  and $F=D^{+}$.

The following lemmas are used in the proof of Theorem \ref{theo1}.

\begin{lem}\label{lem3}\cite{ioannou2012robust}
  If $f, \dot{f} \in L_{\infty}$ and $f \in L_{p}$ for some $p \in[1, \infty),$ then $f(t) \rightarrow 0$ as $t \rightarrow \infty$.
\end{lem}

\begin{lem}\label{lem4}\cite{ioannou2012robust}
  If  $\lim_{t \to \infty}\int_{0}^{t}f(\tau)d\tau$ exists and is finite, and $f(t)$ is a uniformly continuous function, then $\lim_{t \to \infty} f(t)=0$.
\end{lem}

\begin{theorem} \label{theo1}
Under Assumptions \ref{assump1}-\ref{assump2}, consider the nonlinear system (\ref{nonlinear systems}) with unknown disturbance dynamics (\ref{disturbance dynamic}).   Then, the adaptive law (\ref{adaptive law}) along with the disturbance observer (\ref{observer disturbance}), (\ref{estimate state}) guarantees the convergence of the disturbance estimation error $\tilde{\varepsilon}_{T}$ to zero.
\end{theorem}

\begin{IEEEproof}
Let $\tilde{\varepsilon}= \bar{\varepsilon}-\hat{\bar{\varepsilon}}$.    From (\ref{filtered regressor}) and (\ref{estimate state}) it yields
\begin{equation}\label{error e}
\begin{split}
  \tilde{e}=&\bar{\varepsilon}-\hat{\bar{\varepsilon}}=
  \tilde{\varepsilon}
  \end{split}
\end{equation}
 Based on (\ref{derivation varepsilon form}) and
(\ref{estimate state}), one has
 \begin{equation}\label{error dot}
   \dot{\tilde{\varepsilon}}=
   -a\tilde{\varepsilon}+D\tilde{\varepsilon}_{T}
 \end{equation}
According to (\ref{error observer}), (\ref{error e}), and (\ref{error dot}),  one has
 \begin{equation}\label{error dot1}
   \dot{\tilde{e}}=-a\tilde{e}+
   D\tilde{S}F\bar{\varepsilon}+D\tilde{\rho}_{\Delta}
 \end{equation}
Thus, the system (\ref{error dot1}) can be rewritten as
 \begin{equation}\label{vec error}
   \dot{\tilde{e}}= -a\tilde{e}+(\bar{\varepsilon}^{T}F^{T} \otimes D)\tilde{S}_{vec}+D\tilde{\rho}_{\Delta}
 \end{equation}
 where $\tilde{S}_{vec}$ is a vector obtained by stacking rows of the matrix $\tilde{S}$.

Now, consider the following  Lyapunov function candidate.
\begin{equation}\label{Lyapunov function}
  V=\tilde{e}^{T}\tilde{e}+
  \tilde{S}^{T}_{vec}\Gamma^{-1}\tilde{S}_{vec}
\end{equation}

Using (\ref{vec error}) and the adaptive law
(\ref{adaptive law}), the derivative of
(\ref{Lyapunov function}) yields
\begin{equation}\label{derivation Lyapunov function}
\begin{split}
\dot{V}=&-2a\tilde{e}^{T}\tilde{e}+
2a\tilde{e}^{T}D\tilde{\rho}_{\Delta}+\tilde{e}^{T}
(\bar{\varepsilon}^{T}F^{T} \otimes D)\tilde{S}_{vec}\\&+\tilde{S}^{T}_{vec}
(\bar{\varepsilon}^{T}F^{T} \otimes D)^{T}\tilde{e} \\&-
\tilde{S}^{T}_{vec}(\bar{\varepsilon}^{T}F^{T} \otimes D)^{T}\tilde{e}-\tilde{e}^{T}
(\bar{\varepsilon}^{T}F^{T} \otimes D)\tilde{S}_{vec}\\=&-2a\tilde{e}^{T}\tilde{e}+
2a\tilde{e}^{T}D\tilde{\rho}_{\Delta}
\\ \leq & 2a(-\tilde{e}^{T}\tilde{e}
+\frac{1}{2} \tilde{e}^{T}\tilde{e}+2 \|D\|^{2}
\| \tilde{\rho}_{\Delta} \|^{2})
\\ \leq & -a\tilde{e}^{T}\tilde{e}+4a \|D\|^{2}
\| \tilde{\rho}_{\Delta} \|^{2}
%\\ \leq & 4a \|D\|^{2}
%\| \tilde{\rho}_{\Delta} \|^{2}
\end{split}
\end{equation}
Note that based on  (\ref{filtered disrurbance})  and (\ref{observer disturbance}), one has  $\tilde{\rho}_{\Delta}=(\varepsilon_{T}(0)
-\hat{\varepsilon}_{T}(0))e^{-at}$. Since $\tilde{\rho}_{\Delta}$ goes to zero exponentially fast, according to \cite{altafini2019system}, for any $V(t_{e})>0$, there exists a $t_{1}\geq t_{e}$ such that $\forall$ \ $t\geq t_{1}$, $\dot{V} \leq 0$, this implies that $V(t)$ is bounded. Then,  one knows that $\tilde{e} \in L_{\infty}$ and $\tilde{S}_{vec} \in L_{\infty}$.
%Therefore, for any $t \geq 0$

 Furthermore, by integrating (\ref{derivation Lyapunov function}) from both sides, one has
\begin{equation}
\begin{split}
a \int^{t}_{0} \tilde{e}^{T}\tilde{e} d t\leq &-\int^{t}_{0} \dot{V} d t+\int_{0}^{t} 4a \|D\|^{2}\|\tilde{\rho}_{\Delta}(\tau) \|^{2} d \tau\\\leq &V(0)-V(t)+\int_{0}^{t}4a \|D\|^{2}\|\tilde{\rho}_{\Delta}(\tau) \|^{2} d \tau
\end{split}
\end{equation}
The last integral is bounded since $\tilde{\rho}_{\Delta}$  goes to zero exponentially fast, which implies $a \int  \tilde{e}^{T}\tilde{e} d t<\infty $, thus,  $\tilde{e} \in L_{2}$.  Based on Assumption \ref{assump2},  $\varepsilon_{T}$ $\in L_{\infty}$. Then,  using (\ref{disturance with D}) and
(\ref{derivation varepsilon form}), one has $\bar{\varepsilon}$ $\in L_{\infty}$. From $\tilde{e}$  $\in L_{\infty}$ and $\bar{\varepsilon}$ $\in L_{\infty}$, (\ref{error dot1}) concludes that $\dot{\tilde{e}}$  $\in L_{\infty}$,  which together with $\tilde{e} \in L_{2}$ and Lemma \ref{lem3} implies  $\tilde{e}\rightarrow 0$  as $t\rightarrow \infty$. Note that $\hat{\varepsilon} \rightarrow \varepsilon$  as $t\rightarrow \infty$, because $\tilde{e}=\varepsilon-\hat{\varepsilon} \rightarrow 0$  as $t\rightarrow \infty$.  Note also that $\varepsilon=\int_{0}^{t}e^{-a(t-\tau)}\varepsilon_{T}(\tau) d \tau$ and $\hat{\varepsilon}=
 \int_{0}^{t}e^{-a(t-\tau)}\hat{\varepsilon}_{T}(\tau)d \tau$.  Then, $\lim_{t \to \infty}(\varepsilon-\hat{\varepsilon})=
 \lim_{t \to \infty}\int_{0}^{t}e^{-a(t-\tau)}
 (\varepsilon_{T}-\hat{\varepsilon}_{T}) d\tau$=0. From Lemma \ref{lem4}, $\tilde{\varepsilon}_{T}=\varepsilon_{T}-\hat{\varepsilon}_{T} \rightarrow 0$ as $t\rightarrow \infty$.

Therefore, the disturbance estimation error $\tilde{\varepsilon}_{T}$  converges to zero.
This completes the proof.
\end{IEEEproof}

\begin{remark}\label{rem2}
Note that although Theorem \ref{theo1} shows that
 $\tilde{\varepsilon}_{T} \rightarrow 0$, it cannot
 guarantee that $\hat{S} \rightarrow S$. An experience-replay based
adaptive disturbance observer is designed next to estimate $S$ accurately and make the convergence speed much faster.
\end{remark}

\section{Experience-replay based adaptive disturbance observer}
Inspired by \cite{modares2013adaptive, parikh2019integral}, \cite{vahidi2020memory} which used the experience replay for system identification, the experience-replay technique is used to improve the convergence speed of the disturbance observer.  Note that the term $\rho_{\Delta}(t)$ goes to zero exponentially fast; therefore, one can choose a large enough $a$ such that after a short time $t_{0}$, the impact of $\rho_{\Delta}(t)$ is
ignored. The experience replay stores past data in a history stack and reuse them in the disturbance estimation law as
  \begin{equation}\label{adaptive law experence replay x}
  \begin{split}
  \dot{\hat{S}}_{vec}=&\Gamma(\bar{\varepsilon}^{T}F^{T} \otimes D)^{T}\tilde{e}\\&+\kappa \Gamma \sum_{i=1}^{n}Y^{T}_{i}(x(t_{i})-x(t_{i}-\Delta t)-\mathcal{Z}_{i}-\bar{Y}_{i}-Y_{i}\hat{S}_{vec})
  \end{split}
\end{equation}
where $\hat{S}_{vec}$ and $\tilde{S}_{vec}$ are obtained by stacking rows of the unknown matrix $S$ and $\tilde{S}$, respectively, $\Delta t$ is a positive constant denoting the size of the window of integration, $\kappa \in \mathbb{R}^{n}$  is a constant, and $\Gamma$  {\color{black} is a positive  define gain matrix. $t_{i} \in [t_{0}, t]$ are the time points which are between the}  $t_{0}$  and the current time, $Y_{i}=Y(t_{i})$, $\bar{Y}_{i}=\bar{Y}(t_{i})$, and $\mathcal{Z}_{i}=\mathcal{Z}(t_{i})$.
\begin{equation}\label{Y exprience}
Y(t)=\begin{cases}
0,  & \ t \in [t_{0}, t_{0}+\Delta t]\\
\int_{t-\Delta t}^{t} (\bar{\varepsilon}(\tau)^{T} F^{T} \otimes D)d\tau, & \ t> t_{0}+\Delta t\\
\end{cases}
\end{equation}
\begin{equation}\label{Ybar exprience}
\bar{Y}(t)=\begin{cases}
0,  & \  \ \ t \in [t_{0}, t_{0}+\Delta t]\\a
\int_{t-\Delta t}^{t}\bar{\varepsilon}(\tau)d\tau, &  \ \  \ t> t_{0}+\Delta t\\
\end{cases}
\end{equation}
%\begin{equation}\label{rbar exprience}
%\bar{Y}_{1}(t)=\begin{cases}
%0,  & \  \ \ t \in [0, \Delta t]\\D
%\int_{t-\Delta t}^{t}\rho_{\Delta}(\tau)d\tau, &  \ \  \ t> \Delta t\\
%\end{cases}
%\end{equation}

\begin{equation}\label{Z exprience}
\mathcal{Z}(t)=\begin{cases}
0,  & \ t \in [t_{0}, t_{0}+\Delta t]\\
\int_{t-\Delta t}^{t}\phi^{*}z(x(\tau),u(\tau))d\tau , & \ t> t_{0}+\Delta t\\
\end{cases}
\end{equation}

 For any $t>t_{0}+\Delta t$, integrating (\ref{re nonloear system}) yields

 \begin{equation}\label{integrating term}
  \begin{split}
  \int_{t-\Delta t}^{t}\dot{x}(\tau)d\tau=&\int_{t-\Delta t}^{t}
  \phi^{*}z(x(\tau),u(\tau))d\tau \\&+D\int_{t-\Delta t}^{t}\varepsilon_{T}(\tau)d\tau
  \end{split}
\end{equation}

Using (\ref{Y exprience})-(\ref{integrating term}), one has
\begin{equation}\label{calution term}
  \begin{split}
 x(t)-x(t-\Delta t)= Y(t)S_{vec}+\mathcal{Z}(t)+\bar{Y}(t)
  \end{split}
\end{equation}
where $S_{vec}$ is the stacking rows of the unknown matrix $S$.

Substituting (\ref{calution term}) into (\ref{adaptive law experence replay x}) yields

\begin{equation}\label{adaptive law experence replay}
  \dot{\hat{S}}_{vec}=\Gamma(\bar{\varepsilon}^{T}F^{T} \otimes D)^{T}\tilde{e}+\kappa \Gamma \sum_{i=1}^{n}Y^{T}_{i}Y_{i}\tilde{S}_{vec}
\end{equation}

From the adaptive law  (\ref{adaptive law experence replay x}) and (\ref{adaptive law experence replay}), the time is divided into two phases. In the initial phase, {\color{black}the collected data is insufficient to} satisfy a richness condition on the history stack. After a finite period of time, the {\color{black} observer switches to the second phase, where the history stack is sufficiently rich. To assure that the observer switches to the second phase  in finite time,}  sufficiently rich data are required to {\color{black} be  collected} after a  finite period of time {\color{black} as discussed} in the following assumption.
\begin{assump}\label{assump3}
  The system (\ref{nonlinear systems}) is sufficiently excited over a finite duration of time. Specifically, there exist a positive constant $
  \omega $ and time $T>t_{0}+ \Delta t$ for any
  $t> T$, such that $\lambda_{min}(\sum_{i=1}^{n}Y^{T}_{i}Y_{i})>\omega $.
 \end{assump}

 \begin{remark}\label{rem3}
 Compared to the adaptive law (\ref{adaptive law}), (\ref{adaptive law experence replay}) has  an
 extra term which depends on the history of data
collected over time.
 \end{remark}

\begin{theorem} \label{theo2}
Under Assumptions \ref{assump1}-\ref{assump3}, consider the nonlinear system (\ref{nonlinear systems}) with the unknown disturbance (\ref{disturbance dynamic}).  Then, the adaptive control law (\ref{adaptive law experence replay}) along with the disturbance observer (\ref{observer disturbance}), (\ref{estimate state}) guarantee that the unknown dynamic matrix estimation error  $\tilde{S}$ and estimation error $\tilde{\varepsilon}_{T}$ converge to zero  exponentially fast.
\end{theorem}

 \begin{IEEEproof}
Consider the following  Lyapunov function candidate
\begin{equation}\label{Lyapunov function 1}
  V=\tilde{e}^{T}\tilde{e}+
  \tilde{S}^{T}_{vec}\Gamma^{-1}\tilde{S}_{vec}
\end{equation}

Under  Assumption \ref{assump3}, the system (\ref{nonlinear systems}) only requires to be exciting up to time $T$, after which the exciting data recorded during $t \in [t_{0},T]$ is used for all $t>T$.

Then, using (\ref{vec error}) and the adaptive law
(\ref{adaptive law experence replay}),
   during
$t \in [T, \infty)$,  the derivative of (\ref{Lyapunov function 1}) yields
\begin{equation}\label{derivation Lyapunov1}
\begin{split}
\dot{V}=&-2a\tilde{e}^{T}\tilde{e}+\tilde{e}^{T}
(\bar{\varepsilon}^{T}F^{T} \otimes D)\tilde{S}_{vec} \\& +\tilde{S}^{T}_{vec}
(\bar{\varepsilon}^{T}F^{T} \otimes D)^{T}\tilde{e} - \tilde{S}^{T}_{vec}(\bar{\varepsilon}^{T}F^{T} \otimes D)^{T}\tilde{e} \\& -\tilde{e}^{T}
(\bar{\varepsilon}^{T}F^{T} \otimes D)\tilde{S}_{vec}- 2\kappa\tilde{S}^{T}_{vec}\sum_{i=1}^{n}
Y^{T}_{i}Y_{i}\tilde{S}_{vec}
\\=&-2a\tilde{e}^{T}\tilde{e}- 2\kappa\tilde{S}^{T}_{vec}\sum_{i=1}^{n}
Y^{T}_{i}Y_{i}\tilde{S}_{vec}
\end{split}
\end{equation}

According to Assumption \ref{assump3}, $\lambda_{min}(\sum_{i=1}^{n}Y^{T}_{i}Y_{i})>0 $ for any $t \in [T, \infty)$. This implies that $\sum_{i=1}^{n}Y^{T}_{i}Y_{i}$ is positive.

Let $\eta(t)=[\tilde{e}^{T}, \tilde{S}^{T}_{vec}]^{T}$. From (\ref{derivation Lyapunov1}) one has
\begin{equation}\label{derivation Lyapunov2}
\begin{split}
\eta(t) \leq \sqrt{\frac{\varpi_{1}}{\varpi_{2}}}\| \eta(T)\|exp(-\lambda_{1}(t-T))
\end{split}
\end{equation}
where $\varpi_{1}=max\{1, \lambda_{max}(\Gamma^{-1})\}$ and
$\varpi_{2}=min\{1, \lambda_{min}(\Gamma^{-1})\}$.
 $\lambda_{1}=\frac{2}{\varpi_{1}}min\{a, \kappa\omega \}$. Thus, the error $\tilde{e}$ and the estimation error $\tilde{S}$ converge to zero exponentially fast.  Note  that   $\hat{\varepsilon}$ goes to $\varepsilon$  exponentially fast because $\tilde{e}=\varepsilon-\hat{\varepsilon}$ goes to zero exponentially fast. Thus, we can obtain $\tilde{\varepsilon}_{T}=\varepsilon_{T}-\hat{\varepsilon}_{T} \rightarrow 0$ exponentially fast. This completes the proof.
\end{IEEEproof}

\begin{remark}\label{rem4}
Condition (\ref{derivation Lyapunov2}) shows that
 the convergence rate depends on $\omega$ and $a$. Using
an appropriate data selection algorithm for adding  new
samples to the history stack and removing the old ones to increase the minimum eigenvalue of $\sum_{i=1}^{n}
Y^{T}_{i}Y_{i}$ can significant improve the convergence speed.
\end{remark}
{\color{black}
\begin{remark}\label{remk experience}
In \cite{modares2013adaptive, parikh2019integral}, the experience reply is used to  estimate  the identifier weights matrix for a
parameterized nonlinear system. By contrast, this paper leverages the experience replay to estimate the disturbance, which requires new developments.

\end{remark}}
\section{Adaptive Finite time Control law design and stability analysis}

In this section, a finite-time disturbance rejection controller is presented by incorporating the integral terminal sliding mode control (ITSMC) with the proposed disturbance observer. Using Theorem \ref{theo2}, the variation of tendency of disturbance is known and will be leveraged in the control design; therefore, to guarantee the stabilization of the system, the controller's gain does not need to set to a high value in contrast to \cite{feng2019integral, zhu2018adaptive}.

Let define $x_{d}$ as the reference trajectory of the system (\ref{nonlinear systems}) and assume that $\dot{x}_{d}$ is available for the control purpose. Thus, the tracking error is defined as
 \begin{equation}\label{tracking error e}
   e_{x}=x-x_{d}
 \end{equation}

To develop the ITSMC, the sliding surface $\sigma$ is defined as

 \begin{equation}\label{sliding mode surface}
   \sigma=e_{x}+e_{I}
 \end{equation}
 where $e_{I}=\int_{0}^{t}sign(e_{x}(\tau))d\tau$.

 To reject the disturbance, the following controller is designed as

 \begin{equation}\label{controller adaptive sliding mode}
\begin{array} {ll}
u=g^{+}(x)(-f(x)+ \dot{x}_{d}- \\ \ \ \ \ D\hat{\varepsilon}_{T}
   -sign(e_{x})-k(t)sign(\sigma))
\end{array}
 \end{equation}
 where the adaptive controller's  gain $k(t)$ is defined as
  \begin{equation}\label{k adaptive}
  k(t)=
k_{0}+k_{1}\|\bar{\varepsilon}\|e^{-\lambda_{1}  t}
 \end{equation}
 where $k_{0}$ is a small positive constant, $k_{1}\geq \|F\| \|D\|$, and  $\lambda_{1}$  is a positive value defined in Theorem \ref{theo2}.
% which $t_{s}$ is the stable time of the system.  $u^{*}>k_{0}+k_{1}\|\bar{\varepsilon}
% \|e^{-\mathbb{\omega}t_{s}}$ is a constant.
% $k_{0}$ is a very small constant.

The following lemma is used in the proof of  Theorem 3.

\begin{lem}\label{lem6}\cite{bhat1998continuous}
 Consider the following system
 \begin{equation}\label{lem systems}
   \dot{x}=f(x), \ f(0)=0,  \ \ x \in \mathbb{R}^{n}
 \end{equation}
 Let $V(x)$ be defined as a positive definite continuous function which satisfies
 \begin{equation}\label{ly lemma6}
   \dot{V}(x)+\mathbf{a}_{1}V^{\mathbf{a}_{2}}(x)\leq 0
 \end{equation}
 where $\mathbf{a}_{1}>0$ and $0<\mathbf{a}_{2}<1$.  Thus, $x$ converges to the equilibrium point in  finite time.
\end{lem}

The following theorem presents a finite-time control law for disturbance rejection control using the proposed disturbance observer.
\begin{theorem} \label{theo5}
Under Assumptions \ref{assump1}-\ref{assump3}, consider the nonlinear system (\ref{nonlinear systems}) with the unknown disturbance (\ref{disturbance dynamic}). The control law (\ref{controller adaptive sliding mode}) along with the adaptive disturbance observer  (\ref{observer disturbance}) and  adaptive law (\ref{adaptive law experence replay}) ensures that the tracking error $e_{x}$ converges to zero in finite time.
\end{theorem}

\begin{IEEEproof}
 After collecting rich data (i.e., after $t>T$), we use the experience replay to assure {\color{black} the convergence of disturbance estimation error to zero}.  From (\ref{nonlinear systems}) and  (\ref{controller adaptive sliding mode}),  the derivative of the sliding mode surface $\sigma$  can be given as
\begin{equation}\label{ adaptive derivative surface}
  \dot{\sigma}=-k(t) sign(\sigma)+ (\bar{\varepsilon}^{T}F^{T} \otimes D)\tilde{S}_{vec}
\end{equation}

Consider the following Lyapunov function candidate {\color{black}as}
\begin{equation}\label{sliding mode Lyapunov1}
  V=\sigma^{T}\sigma
\end{equation}

Then, the derivative of (\ref{sliding mode Lyapunov1}) is
 \begin{equation}\label{derivative adaptive sliding mode}
 \begin{split}
   \dot{V}=&-2k(t)|\sigma| +2\sigma(\bar{\varepsilon}^{T}F^{T} \otimes D)\tilde{S}_{vec} \\  \leq & -2(k(t)-\| d_{\Delta}
   \|) \|\sigma\| \\  \leq & -2(k(t)-\| d_{\Delta} \|)V^{\frac{1}{2}}
 \end{split}
 \end{equation}
 where $ \|d_{\Delta}\|$ is the {\color{black}bound} of $(\bar{\varepsilon}^{T}F^{T} \otimes D)\tilde{S}_{vec}$,  and $k(t)$ is designed as (\ref{k adaptive}). From Theorem 2, the disturbance estimation error {\color{black}converges} to zero exponentially fast under Assumption \ref{assump2}. {\color{black} Based on Theorem \ref{theo2}, $\kappa\omega \geq \lambda_{1}$.} Therefore, one has

 \begin{equation}\label{k line}
 k(t)=k_{0}+k_{1}\|\bar{\varepsilon}\|e^{-\lambda_{1} t}> k_{1} \|\bar{\varepsilon}\| e^{-\kappa \omega t} \geq \| d_{\Delta} \|
 \end{equation}
  Substituting  (\ref{k line}) into (\ref{derivative adaptive sliding mode}),  the Lyapunov function candidate  (\ref{sliding mode Lyapunov1}) satisfies the finite-time stability condition  (\ref{ly lemma6}) in Lemma \ref{lem6}.  Therefore, for any  initial condition $\sigma(0)\neq 0$, the system (\ref{nonlinear systems}) reaches the sliding manifold $\sigma(t) = 0$ in  finite time.  Then, using (\ref{sliding mode surface}), one has
 \begin{equation}\label{zero sliding mode surface}
   e_{x}=-\int_{0}^{t}sign(e_{x}(\tau))d\tau
 \end{equation}
 which implies that {\color{black} the system} (\ref{nonlinear systems})  converges  to zero along  $\sigma(0)\neq 0$ in  finite time
 after  the system reaches the sliding manifold $\sigma(t)=0$  in finite time \cite{chiu2012derivative, feng2019integral}.  Therefore, the tracking error $e_{x}$ converges to zero in finite time. This completes the proof.
\end{IEEEproof}

 \begin{remark}\label{rem5}
 If the rich data is not collected at the time of the control design, i.e., the condition of Assumption 3 is not satisfied, the proposed controller (\ref{controller adaptive sliding mode}) can be modified as follows by adding another phase to it to make sure that  before Assumption \ref{assump3} is satisfied, the system remains stable.
 \begin{equation}\label{controller adaptive sliding mode first}
 u=
\left\{
\begin{array} {ll} g^{+}(x)(-f(x)+\dot{x}_{d}-D\hat{\varepsilon}_{T}
   -\hslash e_{x}) \ \  {t \leq T} \\ \\ g^{+}(x)(-f(x)+ \ \dot{x}_{d}\\-D\hat{\varepsilon}_{T}
   -sign(e_{x})-k(t)sign(\sigma))\ \ {t > T}
\end{array}
\right.
 \end{equation}
 where $\hslash$ is a positive constant.

 Before rich data is collected, one has
\begin{equation}\label{tracking error system3}
\begin{split}
  \dot{e}_{x}&=\dot{x}-\dot{x}_{d}=
  -\hslash e_{x}+D\tilde{\varepsilon}_{T}
  \\&=-\hslash e_{x}+(\bar{\varepsilon}^{T}F^{T} \otimes D)\tilde{S}_{vec}+D\tilde{\rho}_{\Delta}
\end{split}
\end{equation}
Consequently, it is clear that the system (\ref{nonlinear systems}) is stable during data collection according to the convergence of disturbance estimation error to zero, and $\tilde{\rho}_{\Delta}$ goes to zero exponentially fast.

\end{remark}

{\color{black}
The schematic of the finite-time disturbance rejection using the experience-replay approach is shown in Fig.1
\begin{figure}[H]
\centerline{\includegraphics[width=3.6in, height=2in]{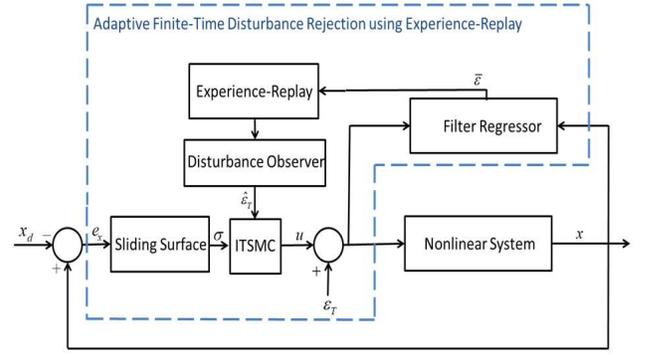}}
\caption{Framework of finite-time disturbance rejection using experience-replay approach.}
\label{fig7}
\end{figure}}

\section{Simulation}
In this section, we present an example to illustrate the effectiveness of the proposed control scheme.

Example 1: Consider the following nonlinear system as
\begin{equation}\label{simulation system}
\begin{aligned}
  \dot{x}_{1}=&x_{1}+x_{2}-x_{1}(x^{2}_{1}+x^{2}_{2}) +u_{1}+\varepsilon_{T_{1}} \\
  \dot{x}_{2}=&-x_{1}+x_{2}-
  x_{2}(x^{2}_{1}+x^{2}_{2})+u_{2}+\varepsilon_{T_{2}}
\end{aligned}
 \end{equation}
 where $x_{1}$ and $x_{2}$ are the states of the system, and $\varepsilon_{T_{1}}$ and $\varepsilon_{T_{2}}$ are the unknown  disturbances.

 Let $z(x,u)=[x_{1} \ \ x_{2} \ \ x_{1}(x^{2}_{1}+x^{2}_{2}) \ \ x_{2}(x^{2}_{1}+x^{2}_{2}) \ \ u]^{T}$, $u=[u_{1} \ \  u_{2}]$, and $\varepsilon_{T}=[\varepsilon_{T_{1}} \ \ \varepsilon_{T_{2}}]^{T}$.  Then, (\ref{simulation system}) can be written as
  \begin{equation}\label{example systems}
  \begin{split}
    \left[
      \begin{array}{c}
        \dot{x}_{1} \\
        \dot{x}_{2} \\
      \end{array}
    \right]
    =& \left[
        \begin{array}{cccccc}
          1 & 1 & -1 & 0 & 1 & 0 \\
          -1 & 1 & 0 & -1 & 0 & 1 \\
        \end{array}
      \right]z(x,u)  \\& +  \left[
                      \begin{array}{cc}
                        1 & 0 \\
                        0 & 1 \\
                      \end{array}
                    \right]\varepsilon_{T}
     \end{split}
  \end{equation}

   The  disturbance dynamics can be expressed as
  \begin{equation}\label{example dynamic}
   \dot{\varepsilon}_{T}=\left[
          \begin{array}{cc}
            0 & \beta \\
            -\beta & 0 \\
          \end{array}
        \right]\varepsilon_{T}=\left[
                                 \begin{array}{cc}
                                   S_{11} & S_{12} \\
                                   S_{21} & S_{22} \\
                                 \end{array}
                               \right]\varepsilon_{T}
  \end{equation}
with $\beta$ as an unknown parameter.

The reference trajectory is $x_{d}=\left[
                                      \begin{array}{c}
                                        2sin 2t \\
                                        4cos 3t \\
                                      \end{array}
                                    \right]$; therefore, the tracking error becomes
   $e_{x}=\left[
    \begin{array}{c}
     x1-2sin 2t \\
     x2-4cos 3t \\
     \end{array}
     \right]$.
{\color{black}
The actual value of $\beta$ is assumed to be $\beta=2$, and the parameters $a$ and $\Gamma$ are selected as  $a=2$ and $\Gamma=50$.

Now,  the adaptive law (\ref{adaptive law})
is used to estimate the disturbance.
Fig. 2 shows that the estimation of the disturbance $ \hat{\varepsilon}_{T}$ goes to the actual disturbance $ \varepsilon_{T}$.
Fig. 3 shows the convergence of the error for all elements of the matrix $S$ in (\ref{example dynamic}).

 Then,  the experience-replay based adaptive law (\ref{adaptive law experence replay}) is use to estimate the disturbance.
Fig. 4 shows that  the estimation of the disturbance $ \hat{\varepsilon}_{T}$ converges to the actual $ \varepsilon_{T}$.  Fig. 5 shows the convergence of the  error  for all elements of the matrix $S$ in (\ref{example dynamic}). Finally, we use the adaptive ITSMC (\ref{controller adaptive sliding mode}) along with the experience-replay based adaptive law (\ref{adaptive law experence replay}) for tracking the reference trajectory $x_{d}$.   Fig. 6 shows that the  tracking errors $e_{x_{1}}=x_{1}-2sin 2t$ and $e_{x_{2}}=x_{2}-4cos 3t$ converge to zero in finite time. }

Comparing Figs. 4-5 to Figs. 2-3, one
 can conclude that the experience-replay based adaptive  observer has much faster convergence speed than the case without using experience replay.

\begin{figure}[H]
\centerline{\includegraphics[width=2.5in, height=2in]{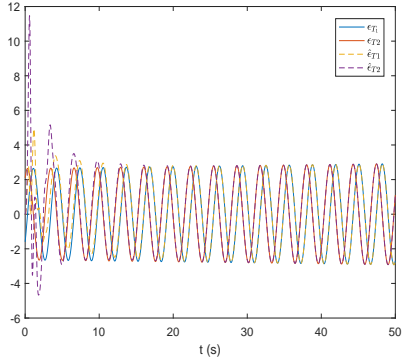}}
\caption{ Disturbance $\varepsilon_{T}$  and the estimation of disturbance $\hat{\varepsilon}_{T}$ without experience replay.}
\label{fig1}
\end{figure}

\begin{figure}[H]
\centerline{\includegraphics[width=2.5in, height=2in]{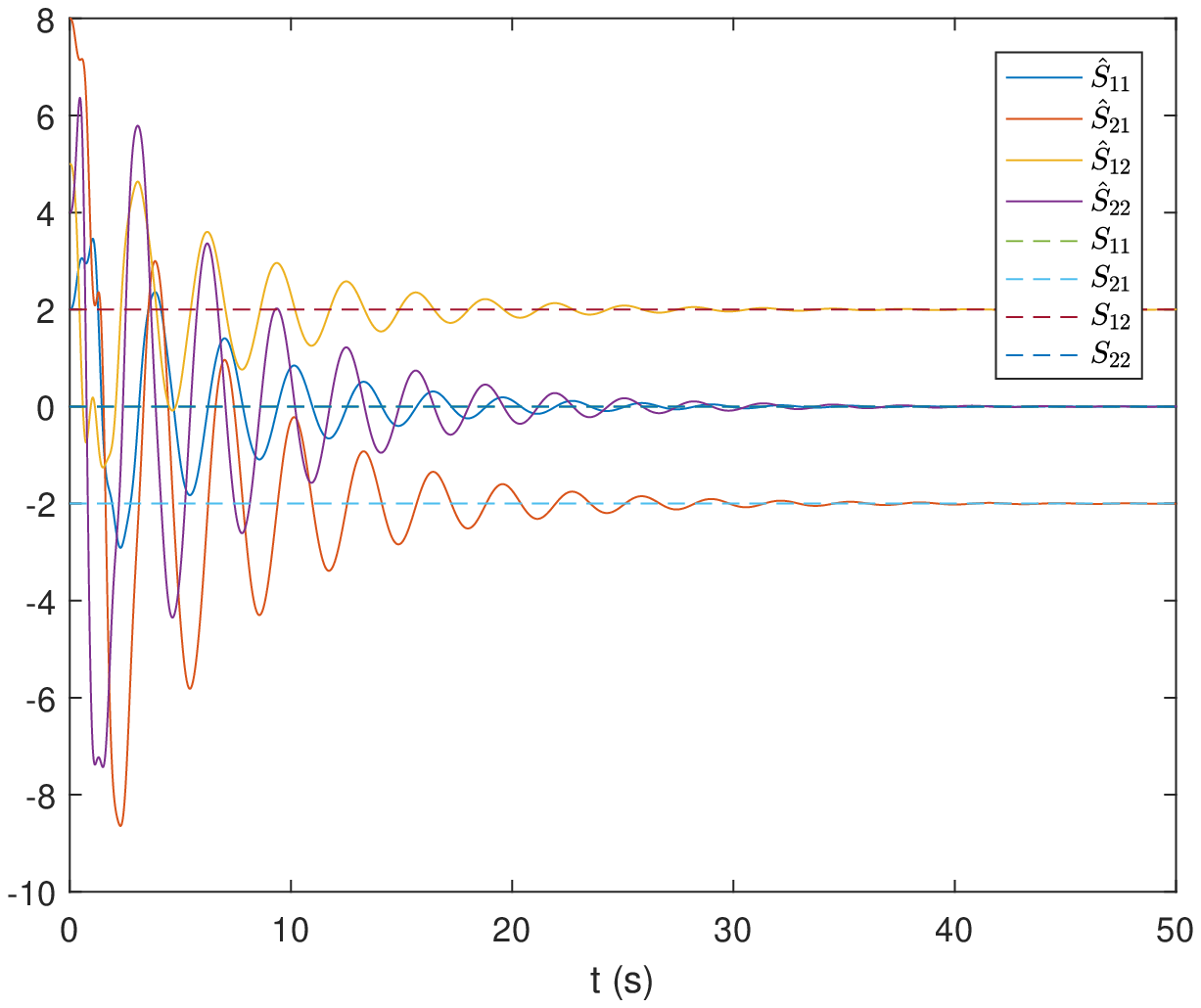}}
\caption{Dynamic matrix elements of $S$ and estimation of dynamic matrix elements of $\hat{S}$ without experience replay.}
\label{fig2}
\end{figure}

\begin{figure}[H]
\centerline{\includegraphics[width=2.5in, height=2in]{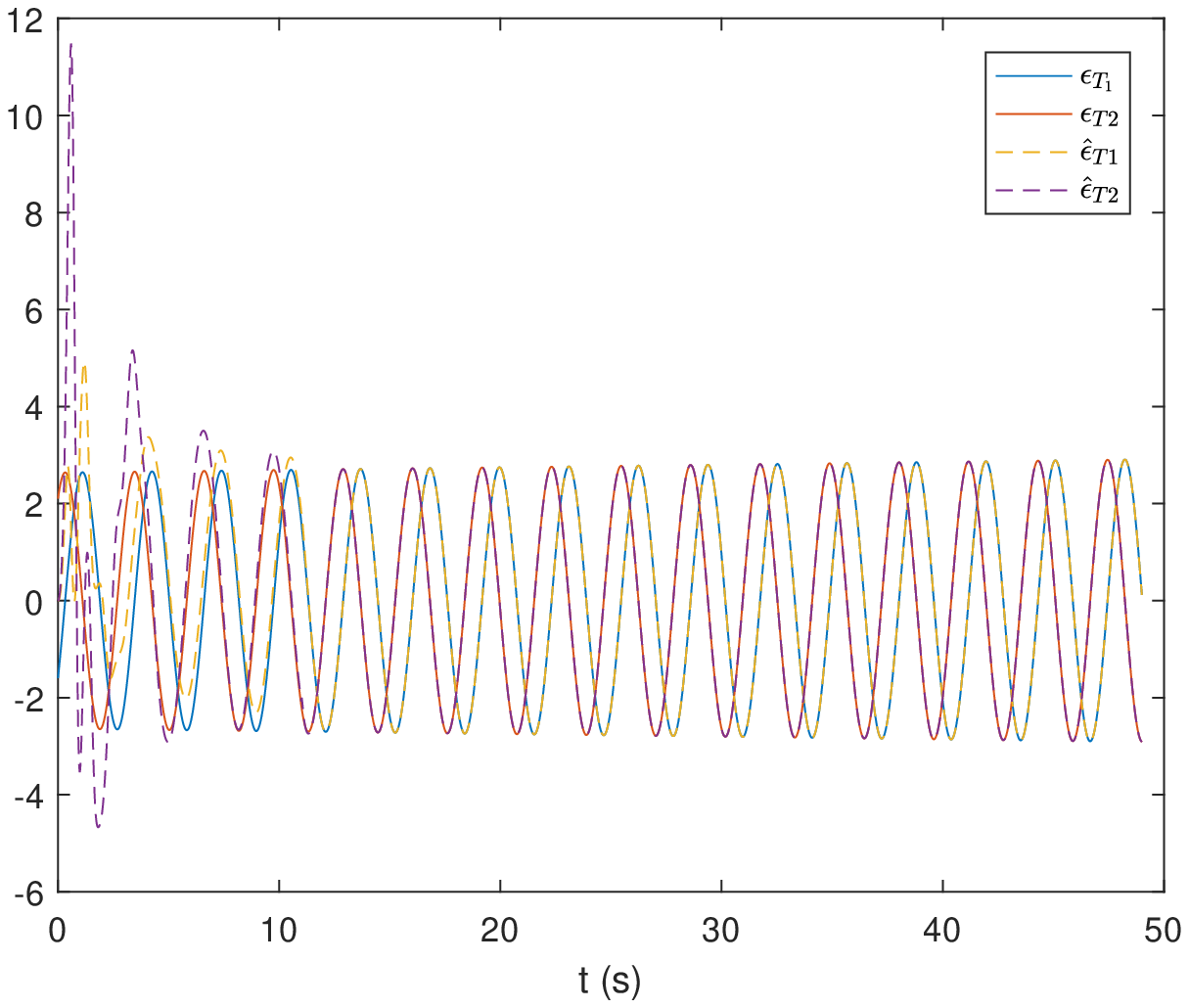}}
\caption{ Disturbance $\varepsilon_{T}$  and the estimation of disturbance $\hat{\varepsilon}_{T}$ with experience replay.}
\label{fig3}
\end{figure}

\begin{figure}[H]
\centerline{\includegraphics[width=2.5in, height=2in]{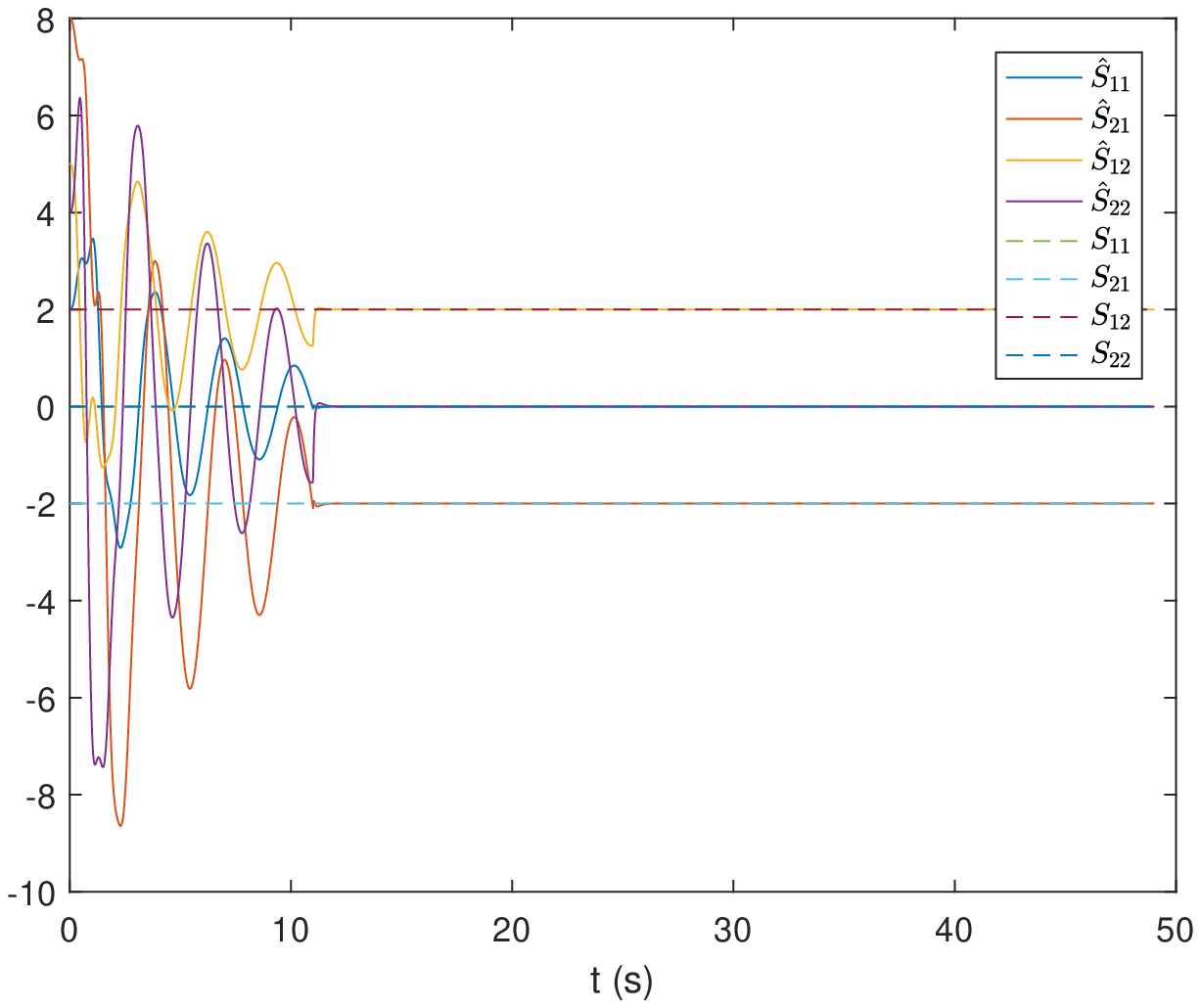}}
\caption{Dynamic matrix elements of $S$ and estimation of dynamic matrix elements of $\hat{S}$ with experience replay.}
\label{fig4}
\end{figure}

\begin{figure}[H]
\centerline{\includegraphics[width=2.5in, height=2in]{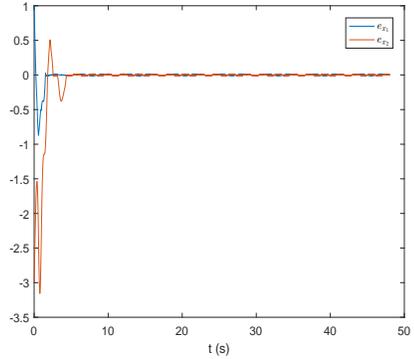}}
\caption{Trajectories of
tracking error $e_{x_{1}}$ and $e_{x_{2}}$.}
\label{fig5}
\end{figure}

%\begin{figure}[H]
%
%\centerline{\includegraphics[width=2.5in, height=2in]{figure/fig6.eps}}
%\caption{Trajectories of
%tracking error $e_{x1}$ and $e_{x2}$.}
%\label{fig6}
%\end{figure}
{\color{black}
Example 2:  Consider the following two-mass-spring system as shown in Fig. 7, which can be used to model a large number of practical systems, including deformable objects' movement and
vibration of mechanical systems
\cite{chen2019reinforcement}. This system is controlled via  $u_{1}$, $u_{2}$    and disturbed by an external force $w$,  where $m_{1}$ and $m_{2}$
denote masses, and $k_{1}$ and $k_{2}$ are spring constants. Defining  $x=[y_{1}, \dot{y}_{1}, y_{2} ,\dot{y}_{2}]^{T}$ as the system state, where  $y_{1}$ and $\dot{y}_{1}$ are the displacement  and velocity of mass $m_{1}$, respectively, $y_{2}$ and $\dot{y}_{2}$ are the displacement of and velocity of mass $m_{2}$, respectively.
Then, the system dynamics with an unknown disturbance are described as
   \begin{equation}
\begin{array}{c}
\dot{x}=Ax+Bu+Dw
\end{array}
   \end{equation}
 where
 \begin{equation}\label{A}
 A=\left[\begin{array}{cccc}
0 & 1 & 0 & 0 \\
\frac{-\left(k_{1}+k_{2}\right)}{m_{1}} & 0 & \frac{k_{2}}{m_{1}} & 0 \\
0 & 0 & 0 & 1 \\
\frac{-k_{2}}{m_{2}} & 0 & \frac{-k_{2}}{m_{2}} & 0
\end{array}\right],
 \end{equation}

 \begin{equation}\label{B}
   B=\left[
   \begin{array}{cccc}
   0 & \frac{1}{m_{1}} & 0 & 0 \\
    0 & 0 & 0 & \frac{1}{m_{2}}  \\
    \end{array}
    \right]^{T},
 \end{equation}
 \begin{equation}\label{D}
   D=\left[
\begin{array}{cccc}
0& 1 & 0 & 1\\
-1& 0 & -1 & 0 \\
\end{array}
\right]^{T}.
 \end{equation}

The dynamics of the unknown disturbance can be expressed as
\begin{equation}\label{example disturbance as}
\dot{w}=\left[\begin{array}{cc}
0 & -\beta \\
\beta & 0
\end{array}\right]w=\left[
\begin{array}{cc}
 S_{11} & S_{12} \\
  S_{21} & S_{22}
   \\
    \end{array}
    \right]\left[
     \begin{array}{c}
      w_{1} \\
       w_{2} \\
       \end{array}
       \right]
\end{equation}
The system parameters are  $m_{1}=1 kg$, $m_{2}=1 kg$, $k_{1}=1 N/s$, and $k_{2}=1 N/s$.

\begin{figure}[H]
\centerline{\includegraphics[width=2.5in, height=1.2in]{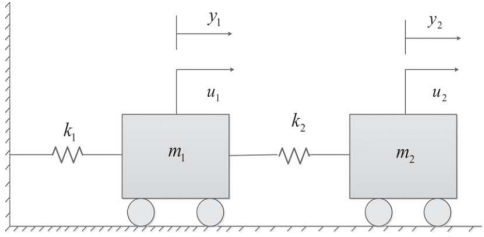}}
\caption{Two-mass-spring system.}
\label{fig6}
\end{figure}
The actual value of $\beta$ is assumed to be $\beta=1.5$, and the parameters $a$ and $\Gamma$ are selected as $a=3$ and $\Gamma=50$.

The reference trajectory is $x_{d}=[sin t \ \ cos t  \ \ cos t \ \ -sin t]^{T} $, and the tracking error is
$e_{x}=x-x_{d} =[e_{x_{1}} \ \ e_{x_{2}} \ \ e_{x_{3}} \ \ e_{x_{4}}]^{T}$.

 Now,  the adaptive law (\ref{adaptive law})
 is used to estimate the disturbance.
Fig. 8 shows that the estimation of the disturbance $ \hat{w}$ goes to the actual disturbance $ w$.
Fig. 9 shows the convergence of the error for all elements of the matrix $S$ in (\ref{example dynamic}).

Then, the experience-replay based adaptive law (\ref{adaptive law experence replay}) is used to estimate the disturbance.
Fig. 10 shows that  the estimation of the disturbance $ \hat{w}$ converges to the actual $ w$.  Fig. 11 shows the convergence of the  error  for all elements of the matrix $S$ in (\ref{example dynamic}). Finally, the adaptive ITSMC (\ref{controller adaptive sliding mode}) along with the experience-replay based disturbance adaptive law (\ref{adaptive law experence replay}) is used for tracking the reference trajectory $x_{d}$.   Fig. 12 shows that the  tracking errors $e_{x_{1}}$, $e_{x_{2}}$ $e_{x_{3}}$, $e_{x_{4}}$ converge to zero in finite time.

Comparing Figs. 10-11 to Figs. 8-9, one
 can conclude that the experience replay based adaptive  observer has much faster convergence speed than the case without using experience replay.

\begin{figure}[H]
\centerline{\includegraphics[width=2.5in, height=2in]{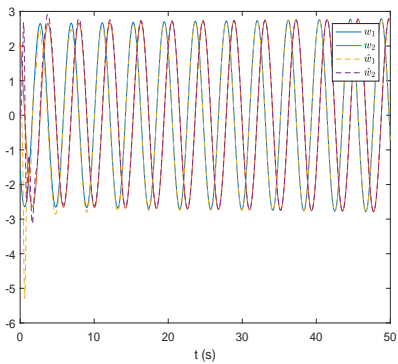}}
\caption{ Disturbance $w$  and the estimation of disturbance $\hat{w}$ without experience replay.}
\label{fig8}
\end{figure}

\begin{figure}[H]
\centerline{\includegraphics[width=2.5in, height=2in]{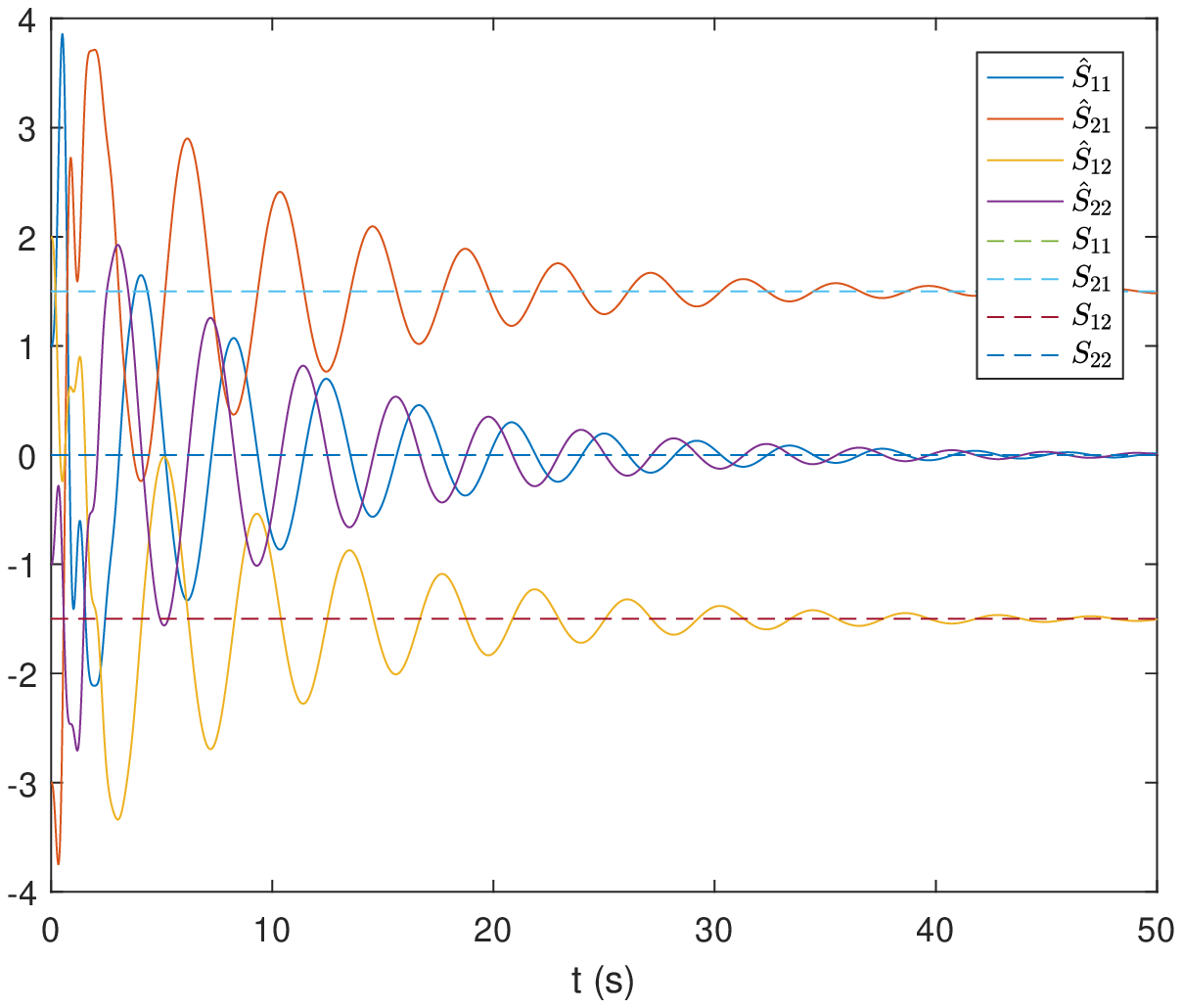}}
\caption{Dynamic matrix elements of $S$ and estimation of dynamic matrix elements of $\hat{S}$ without experience replay.}
\label{fig9}
\end{figure}

\begin{figure}[H]
\centerline{\includegraphics[width=2.5in, height=2in]{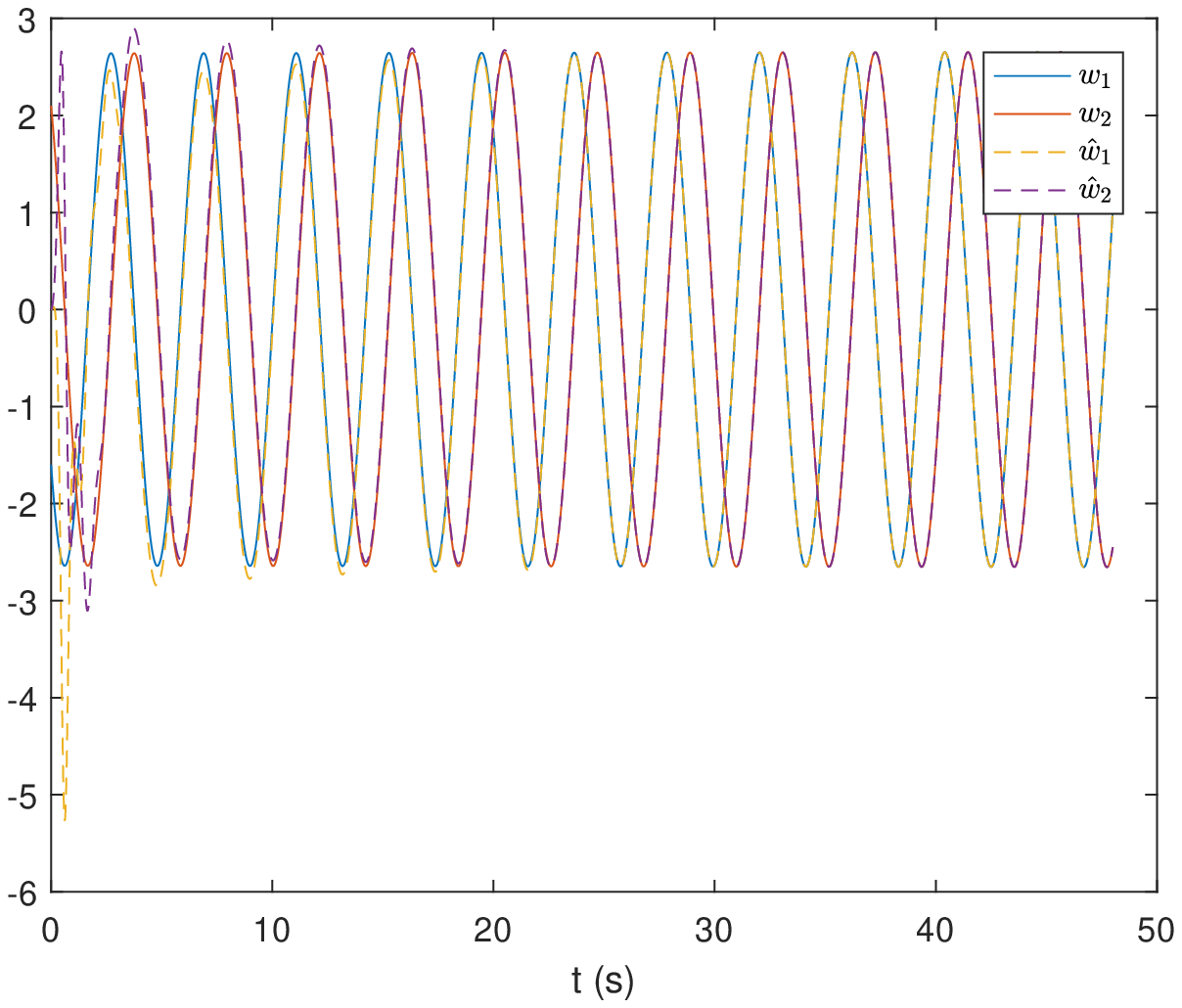}}
\caption{ Disturbance $w$  and the estimation of disturbance $\hat{w}$ with experience replay.}
\label{fig10}
\end{figure}

\begin{figure}[H]
\centerline{\includegraphics[width=2.5in, height=2in]{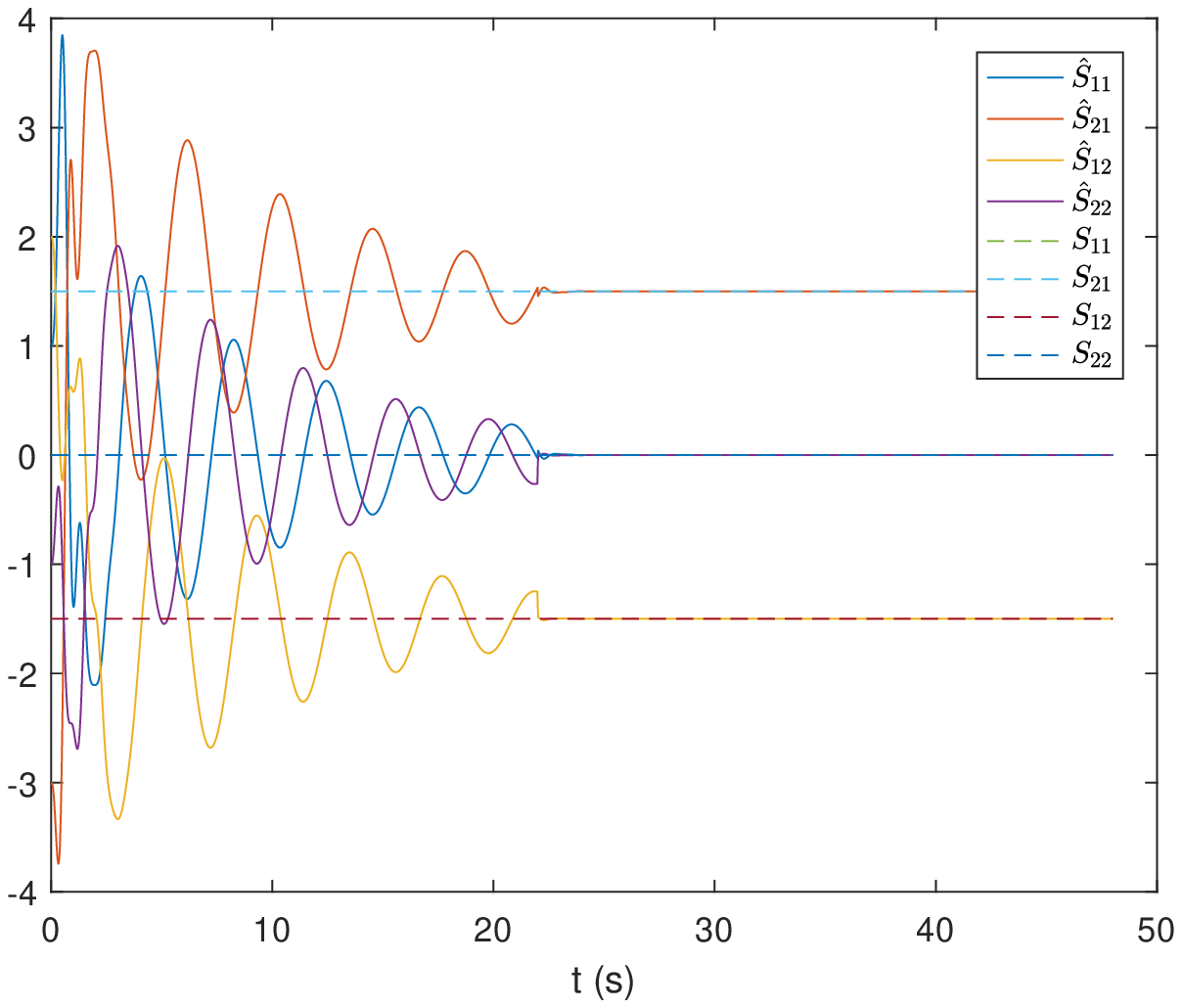}}
\caption{Dynamic matrix elements of $S$ and estimation of dynamic matrix elements of $\hat{S}$ with experience replay.}
\label{fig11}
\end{figure}

\begin{figure}[H]
\centerline{\includegraphics[width=2.5in, height=2in]{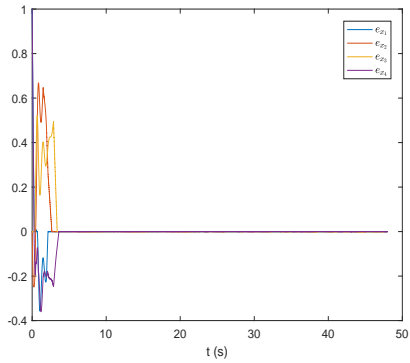}}
\caption{Trajectories of
tracking error $e_{x}$.}
\label{fig12}
\end{figure}

These results confirm that the proposed approach
successfully estimates the disturbance as well
 as it's dynamics,  and the proposed adaptive ITSMC successfully tracks the reference trajectory.
}

\section{Conclusion}
 {\color{black} For a class of systems} with  unknown disturbance,  an adaptive observer was presented  to estimate the disturbance. The proposed
 approach assures that {\color{black}the disturbance estimation error as well as the disturbance exosystem dynamics identification error go to}
 zero exponentially fast.  To achieve this goal,
a filtered regressor form is presented to model both the system dynamics and the disturbance dynamics. This
 allows us to estimate the disturbance  without requiring the measurement of disturbance or state derivatives. Using the experience-replay based {\color{black}adaptive law, convergence} of unknown disturbance dynamics to the actual dynamics is guaranteed. Then, {\color{black} an integral terminal sliding mode controller is presented to assure that the tracking error goes to zero in  finite time. The future work will consider a stochastic framework to take into account the measurement noise and will also consider output feedback control design for disturbance rejection. }

\bibliographystyle{IEEEtran}
\bibliography{IEEEabrv,mybibdisturbancej}

%\linespread{0}
%\fontsize{7.875pt}{0.5mm}\selectfont
%\bibliographystyle{comment}
%\bibliography{mybibdisturbance}

\vspace{12pt}

\end{document}